\documentclass[a4paper,prl,twocolumn]{revtex4}
\usepackage{graphicx}
\begin{document}

\title{Some minor examples on discrete geometry}
\author{Alejandro Rivero}
\affiliation{EUPT (Universidad de Zaragoza), 44003 Teruel, Spain}
\email{arivero@unizar.es}


\begin{abstract}
 By assuming a minimum value for area measurement, the emergence of quantum mechanics 
 can be easily motivated from simple consideration of gravitational
forces. Here we provide some examples and extensions that can be used for pedagogical
purposes. 

At the same time, the role of Planck units is shown to be
of some theoretical influence even at low energies.
\end{abstract}

\maketitle

\section{A Quantum Gravity haiku}

{\it \large Given a particle of mass m, for which radius will a circular 
gravitational orbit around the particle have the property of sweeping 
one Planck area in exactly one Planck time?
}

With this question we began a thread at PhysicsForums website \cite{PF} and
the user Marcus, always optimistic, praised as a "QG haiku".  As a
catching motivation, I argued that below that radius, it should be possible 
to use {\it Planck time beats} to divide area into regions smaller that 
previous. And so, a fundamental break of physics will happen at {\it quantum
Kepler length} of the particle $m$.

Of course this is exaggerated, but 
really the question is about emerging quantum mechanics \cite{ms,s,a1,a2}.
i.e., about to accept the discretizacion of area as a first principle,
then deriving from it the rules of quantum mechanics.
Markopoulou and Smolin \cite{ms} have showed that QM emerges if some
stochasticity principle is incorporated to the description. Here this
small puzzle shows that such principle is a requeriment, as the
answer is $$
R= {\hbar \over mc}$$
ie, Compton radius of the particle $m$. 

We are saved from contradiction because relativistic quantum mechanics
comes to help us: no body can be located beyond its Compton radius.

In 1949, Osborne \cite{osb}, in an unnoticed (except by \cite{misner}) paper,
studied the possibility of measuring the curvature
of a Schwartzchild solution of mass $m$ using geodesical triangles from a test
particle. He applied quantum uncertainty to the test particle and then he
derived sequentially Planck mass, Compton length and Planck length as successive
bounds barring the measurement of curvature. Our example reverses the
path, taking Planck length as the fundamental principle.

The results of \cite{osb} show that, while we have invoked classical gravity for
simplicity, general relativity also contains the same argument.
Also, similar results could surely be derived from other standpoints, for example
the Veneziano \cite[formula 4.2]{triag} string.

\section{Dependence on number of dimensions}
Note that Planck length has dissapeared
above, giving place to the usual QM relationship.
This is a peculiarity
of gravity on 3+1 dimensional space.

Consider a generic force $G mm'/r^q$
so that the units of G will depend on $q$.
In general, asking $A(t_P)$ to be a multiple $n$ of Planck Area $A_P$,
 we have
$$2n=G^{\frac12-\frac1q} m^\frac12 r^{\frac32-\frac{q}2}
c^{\frac3q-1} \hbar^{-\frac1q} $$
And only for the usual inverse-square law, $q=2$, we get to cancel Newton constant.

If we assume that the value of $q$ comes from Fourier transformation of a wave 
propagator, then we are forced to fix space time to be 3+1. 

\section{A dependence on indeterminacy}

But there is also a dimension-independent way to inverse square forces. Time before Feynman,
the mathematician J.L. Singe proposed \cite{singe} to link the potential energy $V(x,t)$ to the total
energy of the photons exchanged to generate $V$. It did not work very well, but it suggests the
following argument:

Assume that the preferent wavelength for exchanged photons is of the order of the distance $r$
between particles, and that this exchange happens under the cloak of indeterminacy principle. 
Then we have a momentum of order
$p \sim \hbar/r$. On the other hand the photon has an energy $E=pc=\hbar c/r$ with the 
associated time $t\sim \hbar/E = r/c$. Thus
$$F={\Delta p\over \Delta t} \sim \hbar c {1 \over r^2}$$

(To justify this $\Delta t$, imagine for instance a stable circular orbit. In this situation the particle changes
momentum but keeps the energy constant. Thus the photon is virtual and it can only exist during
the indeterminacy time. This is the key use of  virtual, off-shell, particles)

It is possible to do the same trick for massive mediators if we start from $\Delta p=\frac 1c \sqrt{
E^2-m^2c^4}$ and, ad hoc, $\Delta t=r/c$. Then one gets a short distance approximation of
yukawian force.

Thus from quantum indeterminacy it seems that forces should be always inverse square. And again
if we want them to come from a wave propagator we are forced to fix space time to be 3+1.

\section{cancellation, or independence}
It could be worth to research the mutual cancellation of the two previous arguments. We could
define gravitational force as the result of a virtual exchange in the above way. Then the units
of G should be independent of space time, and we would always be able
to cancel it and Newton constant to obtain purely the Compton radius. 

In exchange, the Fourier transform of the potential  would  be a "standard wave propagator" 
only for three spatial dimensions.

Regretly this mechanism imposes upon us the need to invoke quantum mechanics, thus it is
muddier than the first, QG only, procedure.

\section{Emergence of quantum mechanics}

 On other hand, if  we have got Compton Length, can we get quantum mechanics from it? 
It is tricky. Compton Length is not exactly a quantum condition, but the 
result of pair creation, via indeterminacy principle, for extreme localisation of energy. We could think
that consistency of quantum gravity implies pair creation and Zitterbewegung, 
but not the whole quantum mechanics. 

Still, we can try in a antique way: the Bohr-Sommerfeld quantum condition can be formulated, at least for circles, via a Newton-Kepler principle: any bound particle sweeps a multiple of Compton Area in a unit of Compton Time. This principle does not need gravity; it works for any central force. Note that we have shifted the point of view; instead
of considering the mass of the central particle, here we have a fixed force
field and we consider the mass of the orbiting particle.

 The usual way to get BS quantisation is to invoke the De Broglie wavelength to check for destructive interference. And then, also, a bound particle sweeps a multiple of De Broglie Area in a unit of De Broglie Time.
 Really if we use any speed $v$ to define area and
time, the same rule apply. While in the first example Planck Length was
cancellated out, here speed simplifies and we are left only
with the quantisation constant.

 A historically minded reader could here enjoy the setup of the area principle
in Newton \cite[book 1, sect 2, Prop 1]{newton}; it is defined first for discrete
areas and impulses.

\section{ugly dimensional analysis}

Naive dimensional analysis can be used also to justify inverse-square forces. In natural units, 
force has a dimension $[L]^{-2}$. In absence of masses, the only scale available is the separation $r$,
between particles. Thus either the force will have a dimensional coupling constant (spoiling 
renormalizability: try naive power counting) or it must use this unique scale available, then 
inducing a dependence on inverse square of distance.

Also, we can use a regulator mass $M$ in adimensional way:
$$
f={K\over r^2}(1-(M r)^p)^q
$$
and then we get a sort of  approximation to short range, yukawian, forces (note q=1/2, p=2 for instance). 

Even if trivial, it is perhaps worth to remark that, when we add some masses, naive dimensional analysis
 offers the possibility of justifying constant and inverse quartic forces. 
The corresponding equations, with a $K$ still adimensional, are
$$
f=K m m' , \mbox{ and }\, f={K\over m m'  x^4}
$$

 This is of some value because a constant force appears as a limit of QCD,
while inverse quartic is a way to approach Fermi theory of contact interactions. The mass in
this later case is known to come from the electroweak bosons $M_W$ and $M_Z$ and not
from the fermions involved.

\section{Scales of mass}

Another user of PF, nicknamed Orion1, suggested to try the two body problem
instead of the Kepler one. I am a bit slow to follow Orion1' calculations, so I have redone them, basically confirming the results. Now, it is interesting to look also to the intermediate steps, so let me play them here.

We have two bodies 1,2 circling around the center of mass, thus with a common angular velocity $\omega$ such that $\omega^2 R_i=G m_j/R^2$. Here R is the sum of both radius. The equation is consistent with the center of mass condition $$R_1m_1=R_2m_2$$

The sum of cases 1 and 2 let us to solve for $\omega$,
$$\omega=\sqrt{G{M\over R^3}}$$

Now we impose that the area $A_i(t_P)$must be a multiple $n_i$ of Plank Area. This translates to
$$n_i=\frac12 \sqrt{cM\over\hbar} {R_i^2\over R^{3/2}}$$

Or, using the C.M. condition to substitute R,
$$R_i={4\hbar\over c} {M^2\over m_j^3} n_i^2$$

Note now that using again this condition over the already solved radiouses, we get a condition on the multiples of area, namely $(m_1/m_2)^2=n_2/n_1$. Or, say, $m_1^2n_1=m_2^2n_2$

Now lets go for the total angular momentum $L=m_1\omega R_1^2+m_2\omega R_2^2$. Substituting and after a little algebra we get
$$L={2 \hbar \over m_P} {M^{3/2}\over (m_1m_2)^{3/2} }
{m_1^7 n_1^4+m_2^7n_2^4 \over(m_1^3n_1^2+m_2^3n_2^2)^{3/2}}$$

Which,  using the relationship between n and m, simplifies to
$$L={2\hbar\over m_P}{m_1+m_2\over m_1m_2} m_1^2 n_1$$

Orion' case $L=\hbar, m_1=m_2\equiv m$ gives us, accordingly, $$m=m_P/4n$$

Also, if we take $m_1$ a lot greater than $m_2$, we recover the initial Compton 
 for $R_2$ and also we get a total angular momentum
$$L_{m_1>>m_2}\approx 2n_1\hbar {m_1^2\over m_P m_2}= 2n_2\hbar {m_2\over m_P}
$$
which shows that Planck mass keeps its role as a bound.

Last, a interesting mistake happens if we try to impose simultaneously low quantum numbers (n1 and n2 small) and big mass differences (m1 a lot greater than m2). Then we are driven to write
$$L^{WRONG}_{m_1>>m_2}\approx  2 n_1 \hbar  { m_1\over m_P} ({m_1\over m_2})^{\frac32}$$
that is not completely out of physical ranges, if for instance we put $m_1$= 175 GeV and $m_2$ of the
order of neutrino differences. The first section of this note has taught us that QFT scales can be implied
by cancelling planckian scales. This last equation, even if unjustified, tell
us that plankian scales can be adequate to study the span of masses in the
known standard model of particles. 
\footnote{for the next step in the ladder, check the plots at \cite{rivero}; the
masses of the standard model, in turn, could generate the extra coupling
that guarantees the scale of nuclear stability.}

\section{ATTACHMENT: Constraints on space-time dimensionality in the classical approximation}

{\it We note the existence, in Newtonian gravity, of two simple arguments to constraint space time dimensionality. They can be used to give space time dimension a values of $D=4$, in the first argument, or $D<5$, in the second case.}

Any theory of gravity will include Newtonian gravity as a limit. Thus it is worth to look there for arguments that in the full theory can be used to restrict the dimensionality of space time, or alternatively to signal a preferred number of non-compactified dimensions. Lets use this column to annotate two arguments of this kind. The calculations are so fast that we will dispense the reader from them.
 
 Theories having discrete units of area and time will meet Kepler's second law in the following way: Consider a test particle orbiting circularly around a body of mass M. Ask for which radius will the particle to sweep a units of Planck area in b units of Planck time. 
 We find that 
 \begin{itemize}
 \item Only for space-time dimension $D=4$ will Newton' constant G cancel out from the calculation. 
 \item In this case, the radius sweeping one unit of Planck area in one unit of Planck time is $R=\hbar /Mc$, the Compton radius of the particle creating the gravitational potential. 
 \end{itemize}
 Those considering quantities such as some density of bound states in a theory will meet Kepler's third law in a peculiar way. Consider two different circular orbits of radius $R_1$, $R_2$ and ask which orbit will a test particle sweep more area for, in the same interval of time. The area being proportional to the square of the radius, the third law tell us that periodes, radiuses and total areas are as $T^2 \sim R^{D-1} \sim A^{(D-1)/2}$, so that for $D=5$ total area is linear with the period of the orbit. Associated to this, we have the following dependence for swept area: 
 
 \begin{itemize}
 \item When $D<5$, it increases when radius increases 
 \item When $D=5$ the area swept by the test particle does not depend of the radius of the orbit, and
 \item When $D>5$, it decreases when the radius of the bound orbit increases 
 \end{itemize}
 
 Of course we should expect that any theory beyond Newtonian gravity will destabilize the criticality of $D=5$, tipping the balance towards one of the two other alternatives.

\end{document}